\documentclass[
  showpacs,
  preprintnumbers,
  reprint,
 amssymb,
 aps,
floatfix,
]{revtex4-1}
\usepackage{graphicx}
\usepackage{bm}
\usepackage{epstopdf}
\usepackage{amssymb}
\usepackage{amsmath}
\usepackage{xcolor}
\usepackage{xfrac}
\usepackage{url}
\usepackage{comment}
\usepackage{hyperref}
\usepackage{cleveref}

\def\XXint#1#2#3{{\setbox0=\hbox{$#1{#2#3}{\int}$}
\vcenter{\hbox{$#2#3$}}\kern-.5\wd0}}

\begin{document}

\preprint{APS/123-QED}

\title{
Irreducible 
nonlocality of optical model potentials based on realistic 
\emph{NN} interactions
}

\author{ H. F. Arellano }
\email{arellano@dfi.uchile.cl}
\affiliation{%
Department of Physics - FCFM, University of Chile,
Av. Blanco Encalada 2008, Santiago, Chile
}
\affiliation{%
CEA,DAM,DIF, F-91297 Arpajon, France
}
\author{ G. Blanchon}
\affiliation{%
CEA,DAM,DIF, F-91297 Arpajon, France
}


\date{\today}

\begin{abstract}
  We investigate the nonlocal structure of optical model potentials
  for nucleon--nucleus scattering based on microscopic approaches. 
  To this purpose, \emph{in-medium} folding optical potentials
  are calculated in momentum space and their corresponding
  coordinate-space counterpart are examined, paying special attention 
  to their nonlocal shape. The nucleon--nucleon effective interaction 
  consists of the actual full off--shell $g$ matrix in 
  Brueckner--Hartree--Fock approximation. 
  The nonlocality of effective interactions is preserved throughout all
  stages in the the calculation. Argonne $v_{18}$ bare potential 
  and chiral next--to--next--to--next--to--leading order bare interaction 
  are used as starting point. The study is focused on proton elastic scattering 
  off $^{40}$Ca at beam energies between 30 and 800~MeV.
  We find that the gradual suppression of high--momentum contributions 
  of the optical 
  potential results in quite different--looking coordinate--space counterparts.
  Despite this non-uniqueness in their nonlocal structure, the implied 
  scattering observables remain unchanged for momentum cutoff above a
  critical one, which depends on incident energy of the projectile. 
  We find that coordinate--space potentials with momentum cutoffs at the
  critical value yield the least structured nonlocal behavior.
  Implications of these findings are discussed.
\end{abstract}

\maketitle
\noindent
\section{Introduction}
The inclusion of nonlocalities in phenomenological optical model
potentials for nucleon--nucleus (\emph{NA}) scattering, 
pioneered in the early 60s by Perey and Buck \cite{Perey1962}, 
made possible simple and robust phenomenological accounts of elastic 
scattering data for targets with nuclear masses in the
range $27\leq A\leq 208$, and nucleon energies of up to 30~MeV.
From a microscopic point of view, considering that the \emph{in-medium} 
density--dependent nucleon--nucleon (\emph{NN}) effective interaction 
is intrinsically nonlocal, it is natural to expect that the optical 
potential itself would turn a nonlocal operator. Moreover, even under 
the simplifying assumption that effective interaction between nucleons 
was local, nonlocal contributions to the optical potential would arise 
from the exchange term after the antisymmetrization of the interaction. 
As we shall demonstrate, for a given optical potential in momentum space 
there is a whole class of potentials in coordinate space with different 
nonlocal structure which yield the same scattering observables. 
The question that comes out then is up to what extent the nonlocal 
features exhibited by microscopic \emph{NA} optical model potentials 
are unique so as to be inferred from \emph{ab-initio} approaches. 

Microscopic models for the optical potential provide us with 
a direct link between the elementary \emph{NN} bare interaction and 
the many--body physics involved in \emph{NA} scattering 
phenomena \cite{Bell1959,Jeukenne1976,Ray1992,Dickhoff2008}.
The fermionic character of all $A+1$ constituents together with
the nonlocal structure of the internucleon effective interaction
result into a nonlocal coupling between single nucleons with the 
remaining $A$ nucleons.
Several efforts have been made in the past decades aimed to 
construct these potentials using realistic \emph{NN} bare
interactions to model effective interactions.
Such is the case of folding models, where a convolution takes 
place between an effective interaction (off--shell $g$ or free $t$ matrix,
depending on the energy of application)
and the ground--state mixed density of 
the target \cite{Ray1992,Amos2000,Aguayo2008,Vorabbi2016}.
Along this line, momentum-- and coordinate--space approaches
adopted by different groups have evolved on parallel tracks,
resulting on comparable description of available scattering data.
In practice, however, specific considerations in the construction and 
treatment of the effective interaction make different momentum-- from 
coordinate--space approaches.
The main difference among them lies in the fact that in the former
the actual fully off--shell $g$ (or $t$) matrix is folded with the 
target mixed density,
whereas in the latter the $g$ matrix is first localized and then folded.
By suppressing nonlocalities in the \emph{NN} effective interaction
(such as Melbourne \cite{Amos2000} or Hamburg \cite{Geramb1983})
nonlocalities in the optical potential arise exclusively 
from the exchange contribution.
This feature contrasts with folding models in momentum space 
\cite{Elster1990,Crespo1990,Arellano1990,Arellano1995,Vorabbi2016},
where both direct and exchange terms result nonlocal.

Another well established microscopic approach for the \emph{NA} optical 
model, particularly at low energies, 
is the one developed by Mahaux and collaborators
\cite{Jeukenne1976}, coined as local density approximation for the
optical potential.
Here the optical potential for nucleon scattering
from a finite nucleus becomes local by construction,
where at each coordinate $r$ of the projectile in the nucleus
it is mapped the on--shell mass operator from infinite nuclear matter,
evaluated at a density equal to that of the target at 
coordinate $r$, namely $\rho(r)$.
The energy at which the mass operator is evaluated corresponds to
that of the beam.
Besides the fact that the resulting optical potential is local,
its spin--orbit part becomes undefined.
This last limitation is fixed with the introduction of a phenomenological 
spin--orbit coupling.
Studies are now being pursued to incorporate a nonlocal absorptive 
contributions to the dispersive optical model within the dispersive 
self--energy method \cite{Dickhoff2017}.

In this work we study the nonlocal structure of microscopic
optical model potentials considering proton scattering off $^{40}$Ca
at beam energies between 30 and 800~MeV. The potentials are calculated 
in momentum space making use of the density--dependent 
Brueckner--Hartree--Fock $g$ matrix \cite{Arellano1995,Aguayo2008}
where its exact nonlocal and off--shell structure are retained.
Once these potentials are transformed into coordinate space
considering different cutoffs at high momenta,
we assess their nonlocal structure as well as equivalence in the
context of scattering observables. We show that the issue of the 
non--uniqueness of the nonlocality is possibly solved considering 
that different looking potentials turn out to share a similar shape 
when a non--reducible momentum cutoff is adopted.

This article is organized as follows. 
In Sec. \ref{ca200} we describe 
our framework and analyze the case of $p+^{40}$Ca elastic 
scattering at 200~MeV. 
We evaluate the potential in momentum space and investigate its structure 
in coordinate space when the high momentum components are suppresses,
with focus on its nonlocal structure.
In Sec. \ref{kdomain} we discuss the existence of a minimum cutoff
applied to optical potentials in momentum space,
above which all scattering observables remain the same.
We explore the nonlocality 
with those of Perey--Buck--type \cite{Tian2015} 
as well as that from a microscopic optical potential 
obtained with the Nuclear Structure Method \cite{Blanchon2015} 
based on Green's function formalism in the 
Random--Phase Approximation using Gogny's effective interaction. 
In Sec. \ref{summary} we summarize and discuss 
the main results of this work.

\section{Proton--nucleus scattering: a study case}
\label{ca200}
Consider the scattering of protons off $^{40}$Ca at beam energy $E$. 
The optical model potential in momentum space can be expressed as the
sum of its central and spin--orbit contributions in the form 
\begin{equation}
\label{omp}
\tilde U({\boldsymbol k}',{\boldsymbol k};E) =
\tilde U_{c}({\boldsymbol k}',{\boldsymbol k};E) +
i {\bm \sigma}\cdot \hat n\;
\tilde U_{so}({\boldsymbol k}',{\boldsymbol k};E)\;,
\end{equation}
with ${\bm \sigma}$ denoting the spin of the projectile and $\hat n$ the unit vector 
perpendicular to the scattering plane defined by 
$\hat n=({\bm k'}\times{\bm k})/||{\bm k'}\times{\bm k}||$. This same operator is 
often denoted as $\tilde\Sigma({\boldsymbol k}',{\boldsymbol k};E)$ 
or $\tilde M({\boldsymbol k}',{\boldsymbol k};E)$ in the literature
\cite{Bell1959,Rotureau2017}.
The optical potential is evaluated in momentum representation 
following procedures outlined in Ref.~\cite{Aguayo2008},
where an \emph{in-medium} effective interaction is folded
with the target full mixed density.
The (nonlocal) density--dependent \emph{NN} effective interaction is 
taken as the actual off--shell $g$ matrix 
solution of the Brueckner--Bethe--Goldstone equation
in the Brueckner--Hartree--Fock approximation.
In the absence of medium effects the $g$ matrix turns the
free $t$ matrix, resulting in the impulse approximation 
for the optical model potential in multiple--scattering 
expansion \cite{Elster1990}.
An appealing attribute of the momentum--space folding approach
we follow is that it enables, within a single framework, 
parameter--free descriptions of nucleon scattering off nuclei 
at energies ranging from few tens of MeV up to 1~GeV
\cite{Arellano1995,Aguayo2008,Arellano2002},
feature not shared by any other reported approach.

To obtain the $g$ matrix we make use of the traditional Argonne 
$v_{18}$ \cite{Wiringa1995} (AV18) bare potential
which has been fit to \emph{NN} phase--shift data at beam energies
below pion production threshold, together with static properties 
of the deuteron. 
We also include in this study a chiral effective--field--theory 
(EFT) interaction, based on chiral perturbation theory. 
The resulting bare interaction is constructed with nucleons and 
pions as degrees of freedom, with the two--nucleon part (2\emph{N}) 
fit to \emph{NN} data. 
We consider the chiral 2\emph{N} force up to 
next--to--next--to--next--to--leading order (N3LO) given 
by Entem and Machleidt \cite{Entem2003}.
To each of these interactions we have calculated their
corresponding infinite nuclear matter single--particle selfconsistent 
fields following Refs.  \cite{Arellano2015,Isaule2016,Arellano2016}, 
to subsequently be used to obtain fully off--shell $g$ matrices.
Radial proton and neutron densities for the $^{40}$Ca target 
are obtained from Hartree--Fock--Bogoliubov calculations based 
on the finite range, density--dependent Gogny's D1S interaction
\cite{Decharge1980}.

\subsection{Momentum-- vs coordinate--space structure}
Typically, momentum--space \emph{NA} potentials are evaluated 
at relative momenta of up to about 10 to 15~fm$^{-1}$
--depending on the beam energy--
choosing an angular mesh $\hat k\cdot\hat k'$ suitable
for reliable partial wave expansions.
Once the central and spin--orbit components of
$\tilde U({\boldsymbol k}',{\boldsymbol k};E)$ are obtained
we extract their corresponding partial wave components $\tilde U_{l}(k',k)$, 
with $l$ the orbital angular momentum.

\begin{figure}
\resizebox{0.80\textwidth}{!}{%
\includegraphics[angle=-0,origin=c,clip=true]{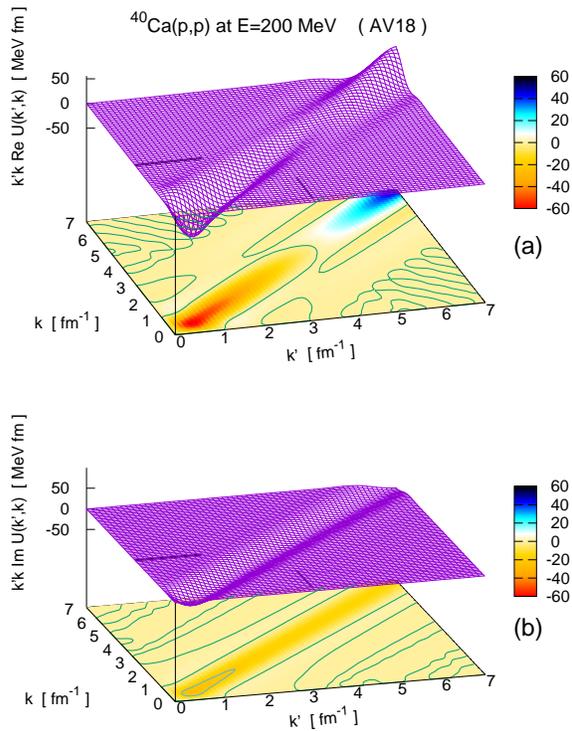}
}
\caption{
  Real (a) and imaginary (b) parts of the
  $s$--wave central optical potential 
  as functions of the relative momenta $k$ and $k'$. 
  Potential corresponding to $p+^{40}$Ca scattering at 200~MeV.
  Results based on AV18 bare interaction.
  Color code in units of MeV~fm.
}
\label{ukk200a}       
\end{figure}

\begin{figure}
\resizebox{0.80\textwidth}{!}{%
\includegraphics[angle=-0,origin=c,clip=true]{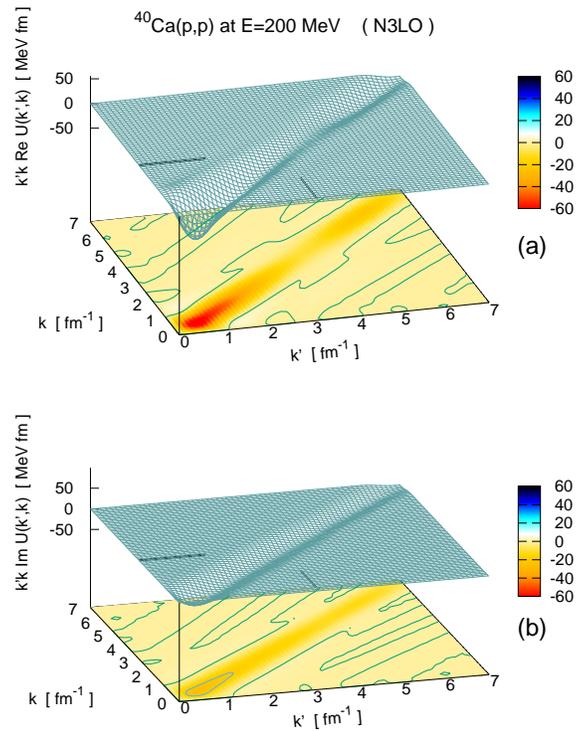}
}
\caption{
  The same as Fig. \ref{ukk200a}, but for N3LO interaction.
}
\label{ukk200n2}       
\end{figure}

Let us first consider $p+^{40}$Ca elastic scattering with proton beam
energy of 200~MeV.
In Figs.~\ref{ukk200a} and \ref{ukk200n2} we show surface plots 
of the real (a) and imaginary (b) $s$--wave components of the 
(central) optical potential, $k'\tilde U_c(k',k)k$,
as functions of the relative momenta $k$ and $k'$. 
Fig.~\ref{ukk200a} and  \ref{ukk200n2} are based on AV18 and 
N3LO bare potentials, respectively.
The short segments on each sheet denote the on--shell momentum.
As observed, 
both real and imaginary components exhibit their dominant contributions 
along a diagonal band. 
In the case of the real component based on AV18, 
a change of sign takes place in the vicinity 
of $k=k'\approx 4$~fm$^{-1}$.
This feature contrasts with that in panel \emph{a} of Fig.~\ref{ukk200n2}, 
where the real part of the potential based on N3LO is weaker at
high momenta, without change of sign.
In the case of the imaginary component, both bare interactions
yield an $s$--wave potential mostly negative maintaining its sign 
along the diagonal.

In coordinate space the central part of the potential reads 
\begin{equation}
\label{q2r}
  U_{l}(r',r) = \frac{2}{\pi}
  \int_0^\infty k'^2dk' 
  \int_0^\infty k^2dk\,
  j_l(k'r') \tilde U_{l}(k',k) j_l(kr)\;,
\end{equation}
resulting a nonlocal function in the general case
(we omit subscript $c$ for simplicity).
A similar expression holds for the spin--orbit term.
Actual evaluations of the above integrals are performed up to some 
upper momenta chosen to guarantee convergence of the evaluated 
scattering observables.
To select the momentum up to which integrals over $k$ are done
we introduce a cutoff function $f_\Lambda(k)$ such that
\begin{equation}
\label{ulambda}
\tilde U(k',k) \to
\tilde U_\Lambda(k',k) =
f_\Lambda(k') \tilde U(k',k) f_\Lambda(k)\;,
\end{equation} 
where 
\begin{equation}
\label{cutoff}
f_\Lambda(k)=
\frac12\left [1-\tanh\left (\frac{k-\Lambda}{\delta}\right )\right ].
\end{equation}
For sufficiently small $\delta$ this function resembles the Heavyside
step function $\Theta(\Lambda - k)$.
Hence, $\Lambda$ defines the momentum above which momenta in
$\tilde U(k',k)$ are suppressed.
Under this definition $\delta$ represents the width of the cutoff,
which for this work we have chosen equal to  $0.2$~fm$^{-1}$ throughout.
This cutoff function plays an analogous role to momentum-space regulators 
in renormalization group transformations \cite{Bogner2010}.

For a given beam energy $E$ we have performed scattering calculations 
in coordinate space considering various values of $\Lambda$,
obtaining practically the same observables when $\Lambda$
lies above certain critical cutoff momentum. 
At the same time, however, different choices of $\Lambda$ above this 
critical momentum lead to potentials with quite different nonlocal structure, 
feature to be addressed in more detail next.

\begin{figure}
 \resizebox{1.0\linewidth}{!}{%
\includegraphics[angle=-90,origin=c,clip=true]{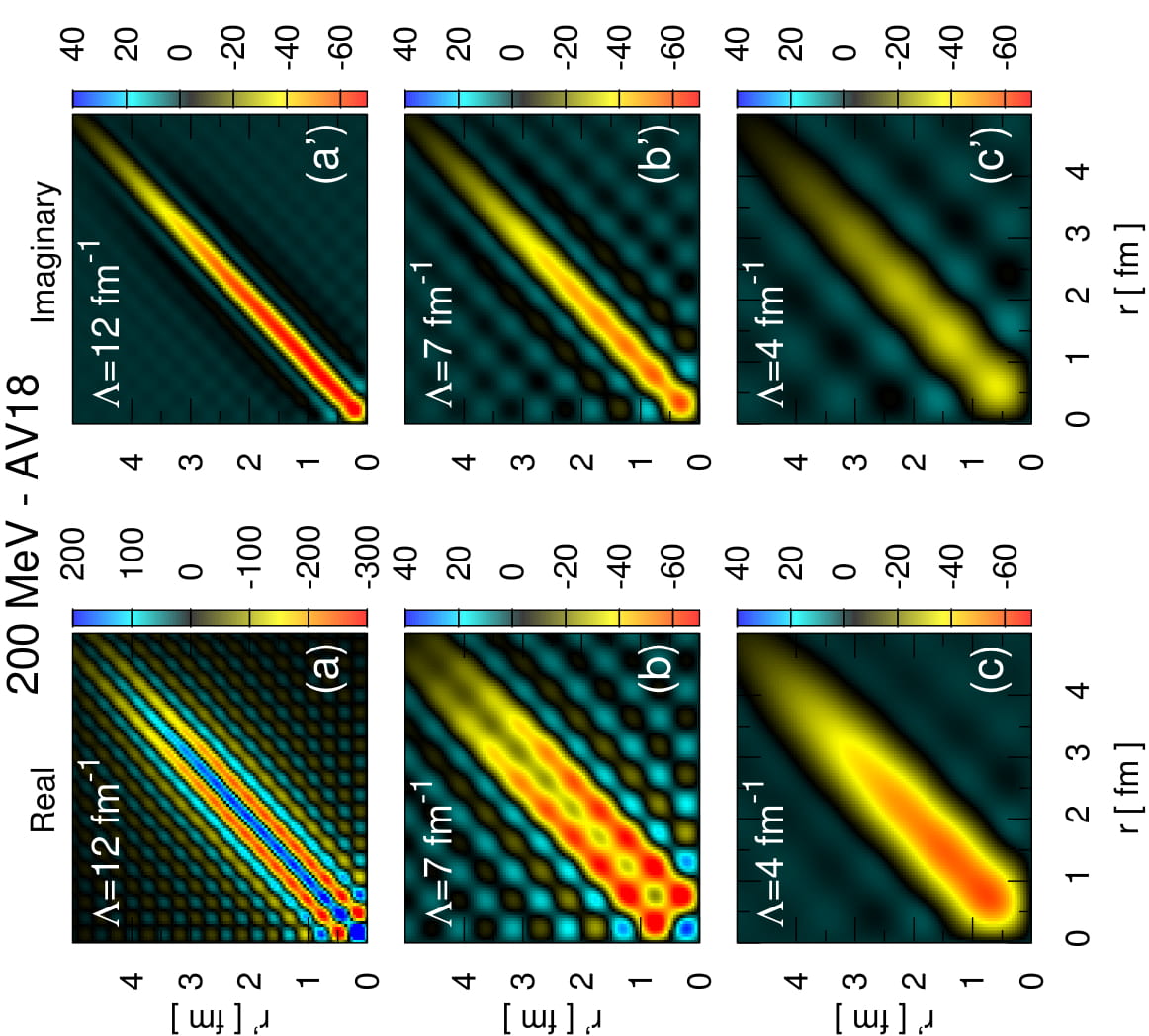}
 }
\caption{
  Contour plots for the real (l.h.s. panels) and
  imaginary (r.h.s. panels) $s$--wave central optical 
  potential as functions of the relative distance $r$ and $r'$. 
  Potential for $p+^{40}$Ca scattering at 200~MeV.
  Results based on AV18 bare interaction.
  Frames a, b and c represent \emph{NA} potentials with cutoffs at
  $\Lambda=12$, $7$ and $4$~fm$^{-1}$, respectively.
  All three potentials yield the same scattering observables.
  Color bar in units of MeV~fm$^{-1}$.
}
\label{urra}       
\end{figure}

\begin{figure}
 \resizebox{1.0\linewidth}{!}{%
\includegraphics[angle=-90,origin=c,clip=true]{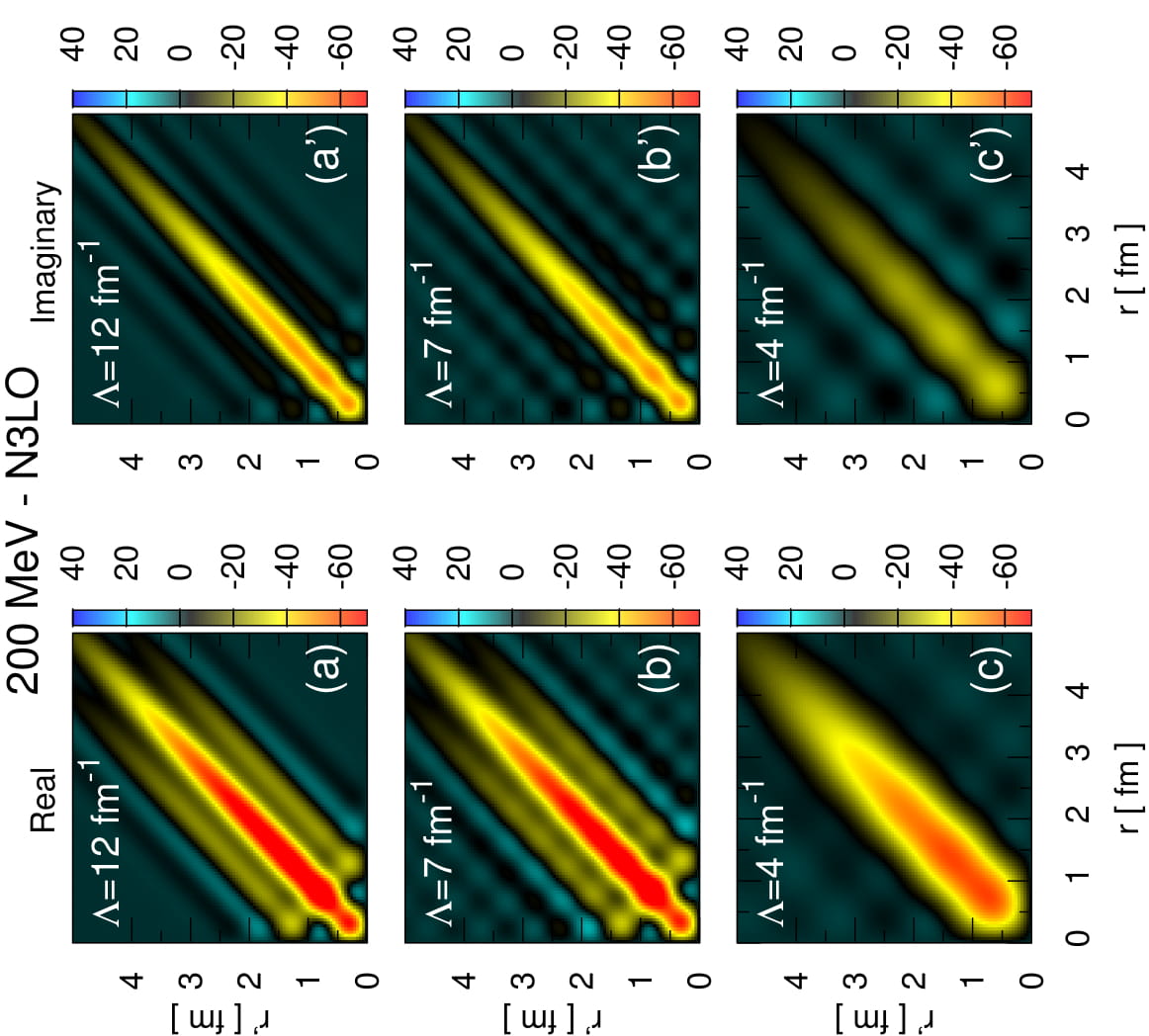}
 }
\caption{
  The same as Fig. \ref{urra}, but for N3LO bare interaction.
}
\label{urrn2}       
\end{figure}

On each of Figs. \ref{urra} and \ref{urrn2} we show six contour plots 
in the $(r,r')$ plane for the calculated $r'U(r',r)r$,
with three choices of $\Lambda$ at and above 4~fm$^{-1}$.
These choices lead to the same scattering observables, 
as will be discussed later.
Figs. \ref{urra} and \ref{urrn2} are based on AV18 
and N3LO bare interactions, respectively.
Panels \emph{a},    \emph{b} and    \emph{c} 
on the left--hand side (l.h.s.) correspond to the real 
component of the potential,
while the prime--labeled panels 
on the right--hand side (r.h.s.) represent the imaginary part.
Panels \emph{a} and \emph{a'} show results for $\Lambda=12$~fm$^{-1}$,
       \emph{b} and \emph{b'}              for $\Lambda=7$~fm$^{-1}$, and
       \emph{c} and \emph{c'}              for $\Lambda=4$~fm$^{-1}$.
What is evidenced in panel \emph{a} of Fig. \ref{urra} is how strong and 
rugged appears
the real part of the AV18--based potential in coordinate space,
as inferred from the narrow diagonal bands of opposite signs
(note the $-300:200$ scale for this panel).
Observe also the narrow oscillatory pattern taking place along 
transverse lines, i.e. those where $(r+r')$ is constant.
The corresponding imaginary component shown in panel \emph{a'} is less
intense, with its sharp dominant contributions located
near the diagonal $r=r'$.
It is worth stating that if the potential was local, 
its plot on any of the panels should result in a narrow 
band along the diagonal $r=r'$.
As the cutoff $\Lambda$ diminishes the potential in coordinate space 
becomes less structured, though it retains its nonlocal nature.
This general feature is evidenced in panels \emph{c} and \emph{c'} of
Figs. \ref{urra} and \ref{urrn2} for $\Lambda=4$~fm$^{-1}$.

\begin{figure}
 \resizebox{0.8\textwidth}{!}{%
\includegraphics[angle=-0,origin=c,clip=true]{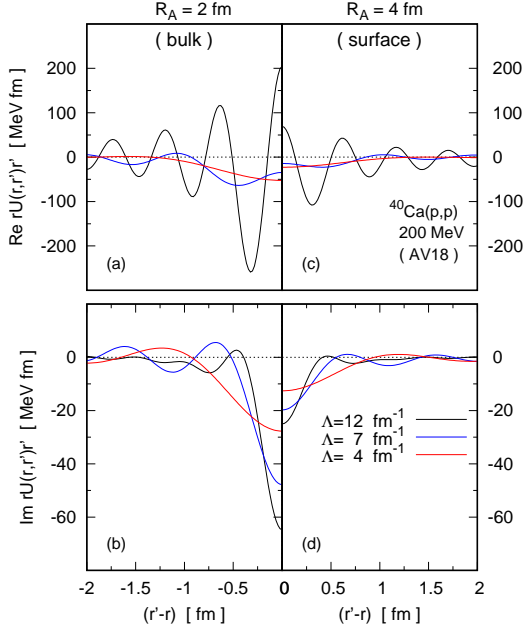}
 }
\vspace{-10mm}
\caption{{\protect\small
\label{trans200a}      
  Nonlocal behavior of the $s$--wave $r'U(r',r)r$ 
  as function of $(r'-r)$, satisfying $r+r'=\textrm{constant}$.
  Black, blue and red curves correspond to results for $\Lambda=12$, 7 
  and 4~fm$^{-1}$, respectively. 
  See text for descriptions of each panel.
}}
\end{figure}

\begin{figure}
 \resizebox{0.8\textwidth}{!}{%
\includegraphics[angle=-0,origin=c,clip=true]{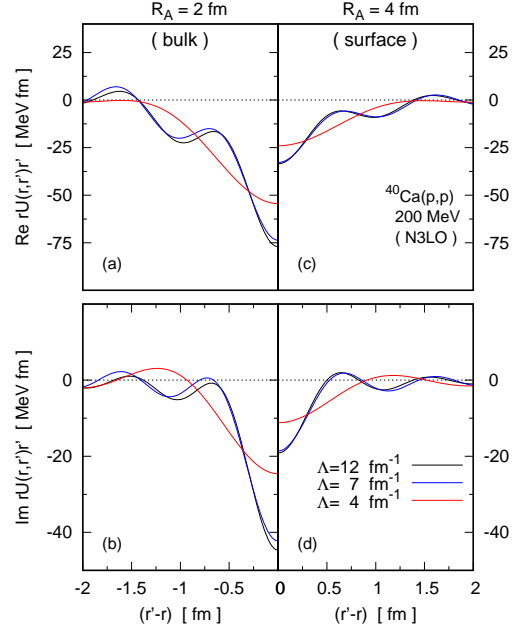}
 }
\vspace{-10mm}
\caption{{\protect\small
\label{trans200n2}     
  The same as Fig. \ref{trans200a}, but for N3LO.
}}
\end{figure}

Clearly the momentum cutoff $\Lambda$
has an incidence on the shape of the potential in coordinate space,
particularly with regard to its nonlocality.
For a closer scrutiny of this feature, in Figs.~\ref{trans200a} (for AV18)
and \ref{trans200n2} (for N3LO) we plot $r'U(r',r)r$ as functions of the 
difference $(r'-r)$ along transverse lines where $r+r'$ is kept constant.
Black, blue and red curves correspond to $\Lambda$ equal 
to 12, 7 and 4~fm$^{-1}$, respectively.
Panels \emph{a} and \emph{b} on the l.h.s. represent $r'U(r',r)r$ 
in the case $(r+r')/2\equiv R_A$, with $R_A=2$~fm.
When $r\approx r'$, this choice of $R_A$ represents a region near the 
bulk of the nucleus.
Panels \emph{c} and \emph{d} on the r.h.s. correspond to 
case $R_A=4$~fm, near the surface of the nucleus.
Upper (lower) panels show the real (imaginary) part of the potential.
We observe that the real part oscillates as a function of $r'-r$,
being stronger for $r=r'$, and decreasing as $|r'-r|$ increases.
The imaginary part behaves similar to the real part, 
but featuring more damped oscillations.
When comparing l.h.s with r.h.s. panels, 
we observe that the potential weakens at the surface.

\begin{figure}
 \resizebox{0.5\textwidth}{!}{%
\includegraphics[angle=-0,origin=c,clip=true]{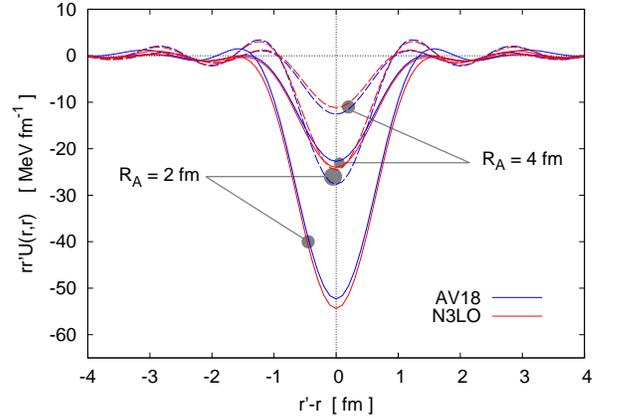}
 }
\vspace{-10mm}
\caption{{\protect\small
\label{nonlocal200}    
Nonlocal behavior of the $s$--wave optical potential at bulk and surface
vicinities in the case of $\Lambda=4$~fm$^{-1}$. 
The optical potential corresponds to $p+^{40}$Ca elastic scattering at 200 MeV
beam energy. 
Blue and red curves denote results based on AV18 and N3LO bare interactions, 
respectively.
}}
\end{figure}

Any near local potential would resemble some representation
of the Dirac $\delta$ function of finite width along transversal lines. 
The curvature at $r'=r$, would serve to gauge the nonlocality
at $R_A=r$. 
From the plots shown in Figs. \ref{trans200a} and \ref{trans200n2} 
we observe a clear sensitivity of the curvature of the potential at $r=r'$, 
under the three values of $\Lambda$, evidencing different 
nonlocalities among them.
What emerges as an interesting property is that as $\Lambda$ 
diminishes, the coordinate--space optical potential based on
AV18 and N3LO become similar, feature we illustrate more closely
in Fig. \ref{nonlocal200} for the $s$--wave component.
Here blue and red curves represent results based on AV18 and N3LO 
bare interactions, respectively. 
Solid (dashed) curves correspond to the real (imaginary) 
component of the potential. 
Labels for $R_A$ identify regions in the nucleus at which the
potential is plotted.
Observe that the real component of the potential is slightly more 
intense for N3LO than for AV18, whereas their corresponding absorptive 
parts are very similar.
Note also that all components feature damped oscillations
as $|r-r'|$ increases. 
The question that arises at this stage is whether differences 
in the nonlocal structure of these potentials, 
as driven by the cutoff $\Lambda$,
have any incidence on the calculated scattering observables.

\subsection{Invariability of scattering observables}
A test of the equivalence of $U_\Lambda$ in coordinate space,
for $\Lambda$ above its critical value,
comes from the actual calculation of scattering observables. 
In this case we solve the integral equation for the scattering
wavefunction
\begin{align}
\label{wave}
  u_{jl}(r) =&
\frac{1}{k} F_l(kr) +
  \left (\frac{2m}{\hbar^2}\right ) \int\int dr'dr'' \nonumber \\
 & G_l^{(+)}(r,r';k) \, 
r'U^{[s]}_{jl}(r',r'') r''\,  u_{jl}(r'') \; ,
\end{align}
leading to exact solutions for the scattering waves in the presence of
Coulomb interaction~\cite{Arellano2018}.
Here $F_l$ denotes the regular Coulomb function for partial wave $l$, 
$j$ the total angular momentum,
and $k$ the momentum of the projectile in the \emph{NA} 
center of momentum reference frame.
The propagator $G_l^{(+)}(r,r';k)$ represents the Green's function 
associated to the Coulomb interaction of point particles of charge 
$Ze$ (nucleus) and $e$ (incident proton) \cite{Arellano2007B}.
Furthermore, $U^{[s]}$ consists of the (nonlocal) optical potential
superposed to the Coulomb--screened interaction, namely
\begin{equation}
  \label{ushort}
r'U^{[s]}_{jl}(r',r) r= r'U_{jl}(r',r)r + 
  \left [
    V_C(r)- \frac{Ze^2}{r} 
  \right ]
\delta(r'-r)\;,
\end{equation}
resulting a nonlocal finite range interaction.
Here $V_C(r)$ represents the electrostatic interaction
between the charge distribution of the nucleus and the charge of the
projectile.
The hadronic component itself is composed by its central and spin--orbit
contributions as 
\begin{equation}
\label{ujl}
U_{jl}=U_{l}^{(c)} + 
\langle{\bm\sigma}\cdot{\bm\ell}\,\rangle_{jl}\; U_{l}^{(so)}\;,
\end{equation}
with
$\langle{\bm\sigma}\cdot{\bm\ell}\,\rangle_{jl}=
[j(j+1)-l(l+1)-\sfrac{3}{4}]$.

\begin{figure}
 \resizebox{0.9\textwidth}{!}{%
\includegraphics[angle=-0,origin=c,clip=true]{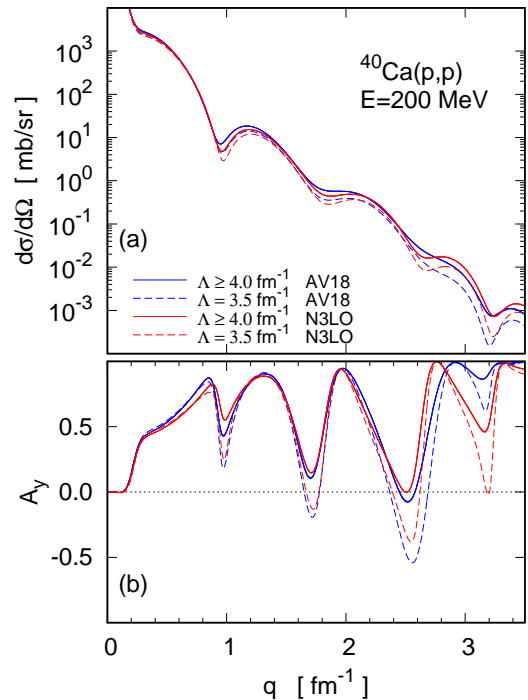}
 }
\vspace{-15mm}
\caption{{\protect\small
\label{xay200}     
  Differential cross section (a) and analyzing power (b) as functions of 
  the momentum transfer for $^{40}$Ca($p$,$p$) scattering at 200~MeV. 
  Blue (red) curves correspond to results based on AV18 (N3LO) 
  bare interaction.
  Solid curves correspond to $\Lambda=12$, 7 and 4~fm$^{-1}$, whereas
  dashed curves correspond to $\Lambda=3.5$~fm$^{-1}$.
}}
\end{figure}

In Fig. \ref{xay200} we show the calculated 
differential cross section ($d\sigma/d\Omega$) and 
analyzing power ($A_y$) for proton elastic scattering from
$^{40}$Ca at 200 MeV beam energy. 
Blue and red curves denote results based on AV18 and N3LO, respectively.
Observables are calculated in coordinate space using 
four values for $\Lambda$.
Three of these correspond to $\Lambda=12$, 7 and 4~fm$^{-1}$, 
whose corresponding observables are all plotted
with solid curves, becoming indistinguishable to the eye.
The case $\Lambda=3.5$~fm$^{-1}$ (dashed curves) has been chosen
deliberately below the critical one in order to contrast its 
results with the other three values.
As observed, all results with $\Lambda\geq 4$~fm$^{-1}$ yield 
practically identical $d\sigma/d\Omega$ and $A_y$.
Such is not the case for $\Lambda=3.5$~fm$^{-1}$, 
resulting in weaker differential cross section and more structured $A_y$.
Regarding the reaction cross section, 
all three cases with $\Lambda\geq 4$~fm$^{-1}$ for AV18 (N3LO)
yield     $\sigma_R=540.0\,(524.9)$~mb, whereas for $\Lambda=3.5$~fm$^{-1}$,
we obtain $\sigma_R=505.0\,(490.8)$~mb. 
Differences of about  35~mb are consistent with the trend 
observed for $d\sigma/d\Omega$ in Fig. \ref{xay200}.

\subsection{Volume integrals}
As a means to check consistency of the results,
we have calculated volume integrals of the central part of the 
optical potential in the two representations under study.
In coordinate space we evaluate explicitly
\begin{equation}
\label{jvol}
J = \int d{\bm r}\, d{\bm r}' U_{c}({\bm r}',{\bm r};E)\;,
\end{equation}
with $U_{c}$ the central component of the optical potential.
Considering that 
\begin{equation}
\label{ukkurr}
\tilde U_{c}({\bm k}',{\bm k};E)=\frac{1}{(2\pi)^3} \int
 d{\bm r}\, d{\bm r}'
e^{-i{\bm k}'\cdot{\bm r}'} 
U_{c}({\bm r}',{\bm r};E)
e^{i{\bm k}\cdot{\bm r}} \;,
\end{equation}
it becomes evident that 
\begin{equation}
\label{jvol2}
J=8\pi^3\,\tilde U_{c}({\bm k}',{\bm k};E)\rvert_{{\bm k}'={\bm k}=0}
\end{equation}
Thus, in momentum space the volume integral $J$ can be obtained from 
the potential matrix element at $k'=k=0$, 
unaffected by any finite cutoff with narrow width.
In the case of the optical potential based on AV18 we obtain
$J/A=-(536.2+ i\,225.9)$~MeV~fm$^{3}$ 
in momentum space,
while in coordinate space we obtain for the four cutoffs $\Lambda$,
$J/A=-(536.7+i\,226.1)$~MeV~fm$^{3}$.
In the case of results based on N3LO we obtain
$J/A=-(571.6+i\,199.8)$~MeV~fm$^{3}$ in momentum space,
while in coordinate space all four cutoffs yield
$J/A=-(572.2+i\,200.0)$~MeV~fm$^{3}$.
The differences among the two representations, close to 0.1\%, 
can be attained to the various numerical procedures involved 
in multipole evaluations and quadratures.

\section{Non--reducible $k$--domain}
\label{kdomain}

We have calculated momentum--space optical model potentials
for proton elastic scattering off $^{40}$Ca 
at several beam energies $E$ between 30 and 800 MeV.
The procedures we follow are the same as the ones applied in 
the previous section.
Considering that realistic \emph{NN} interactions are designed
for nucleon beam energies of up to $\sim$300~MeV, applications 
for $E\geq 400$~MeV are based on procedures described 
in Ref. \cite{Arellano2002}, where minimal relativity corrections 
are included in the evaluation of the optical potential.
The same reference describes the inclusion of a separable term 
added to the bare interaction, in order to reproduce exactly 
empirical \emph{NN} scattering amplitudes above pion production threshold. 
Apart from these considerations, the folding structure of the optical 
potential at these energies are the same as described in Sec.~\ref{ca200}.
For consistency, we have excluded the N3LO bare potential 
from applications at these high energies.

\begin{figure} [ht]
 \resizebox{0.5\textwidth}{!}{%
\includegraphics[width=\linewidth,angle=0,origin=c,clip=true]{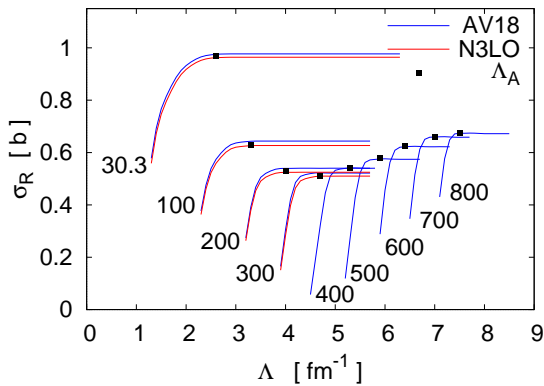}
 }
\vspace{-15mm}
\caption{{\protect\small
\label{reaction}       
Reaction cross section as a function of the cutoff $\Lambda$ applied
to the optical model potential for $p+^{40}$Ca elastic scattering.
Black and red curves represent results based on AV18 and N3LO bare
interactions. Labels on each curve refer to the beam energy. 
Blue squares denote the critical cutoff.
}}
\end{figure}

Once the \emph{NA} potentials are evaluated we proceed to calculate 
their scattering observables with various choices of cutoff $\Lambda$, 
aiming to identify domains that yield the same results within a given 
accuracy.
In Fig. \ref{reaction} where we plot reaction cross sections $\sigma_R$ 
for $p+^{40}$Ca elastic scattering as functions of $\Lambda$. 
Numerical labels indicate the beam energy in MeV units, while
blue and red curves denote results based on AV18 and 
N3LO bare interactions, respectively.
We observe that all curves for $\sigma_R$ as functions of $\Lambda$
reach a plateau above certain critical value, which we denote $\Lambda_A$.
The subscript $A$ is intended to distinguish this cutoff,
applied to \emph{NA} optical potentials,
from those used in renormalization group transformations applied to
\emph{NN} potentials.
Thus, for a given energy, if we apply a cutoff to the momentum--space
potential at or above its corresponding $\Lambda_A$, 
the scattering observables remain unaffected.
This property leads to the notion that there is a minimum cutoff
which renders \emph{NA} scattering observables unchanged.
We shall refer to the complement of that interval as 
non--reducible $k$--domain.
Cutoffs within the non--reducible $k$--domain alter the scattering 
observables associated to the optical potential under consideration.
As we have seen in the previous section, different choices of
$\Lambda$ above $\Lambda_A$ yield different coordinate--space
structure, with the case $\Lambda=\Lambda_A$, resulting in the least 
structured potential.
This feature hints us a means to compare on an equivalent footing
nonlocal structures of different potentials.
Henceforth, potential with cutoff at $\Lambda_A$ shall referred
as reduced potentials, with its implied nonlocal behavior identified
as reduced nonlocality.

\begin{figure}
 \resizebox{0.5\textwidth}{!}{%
\includegraphics[angle=-0,origin=c,clip=true]{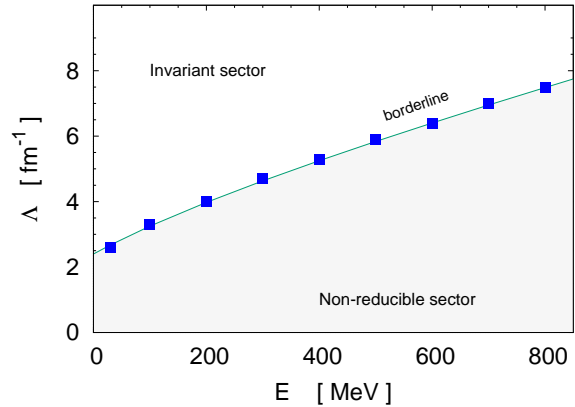}
 }
\vspace{-10mm}
\caption{{\protect\small
\label{lambda}         
Critical cutoff $\Lambda_A$ for the non--reducible  $k$--domain
as function of the beam energy $E$
in the case of proton elastic scattering from $^{40}$Ca.
}}
\end{figure}

Examining more closely Fig. \ref{reaction} for $\sigma_R$ we observe 
that the position of $\Lambda_A$ increases monotonically with the energy.
To obtain $\Lambda_A$ at the edge of the plateau we look for the 
smallest $\Lambda$ which satisfies 
\begin{equation}
\label{plateau}
\left |
\frac{\partial\sigma_R}{\partial\Lambda}
\right | 
\leq\varepsilon\;,
\end{equation}
with $\varepsilon=0.01$~mb~fm.
In Fig. \ref{lambda} we plot with filled squares the obtained 
values of $\Lambda_A$.
We have found that this behavior can be characterized by means of the formula
\begin{equation}
\label{law}
\Lambda_A(E) = \sqrt{\Lambda_0^2 + k_E^2}\;,
\end{equation}
which is plotted with solid curve (borderline).
In this case $\Lambda_0=2.4$~fm$^{-1}$, with $k_E$ the relative momentum 
in the center of momentum reference frame associated to the beam energy $E$.
The region below the borderline corresponds to the non--reducible
$k$--domain, while the region above it represents
the sector which yield scattering observables invariant.
Eq.~(\ref{law}) for $\Lambda_A$ has to be taken as an estimate
of the borderline since it depends on the criteria set for $\varepsilon$,
width $\delta$ of the cutoff as well as numerical accuracy.
This borderline may also depend on the optical model as well as nature
of the \emph{NN} effective interaction under use.
Its extension to proton collisions from targets
other than $^{40}$Ca is beyond the scope of this study.

\subsection{$p+^{40}\!\textrm{Ca}$ scattering
 at 30.3~M\lowercase{e}V}
\label{ca30}
We now study proton scattering from $^{40}$Ca at 30.3 MeV,
an energy which allows us to extend the previous discussion to other 
documented nonlocal optical potentials developed for $p+^{40}$Ca elastic 
scattering \cite{Blanchon2015,Tian2015}.
Apart from the AV18-- and N3LO--based optical potentials, we include 
results based on the microscopic optical model of Ref. \cite{Blanchon2015}
within the NSM
based on Green's function formalism in the Random--Phase 
Approximation using Gogny's effective interaction.
Here we also consider the Perey--Buck--type parametrization
reported by Tian \emph{et al.} \cite{Tian2015}, 
phenomenological approach of common use.

The last two potentials are calculated in coordinate space,
with its equivalent central and spin--orbit 
counterparts in momentum space obtained from
\begin{equation}
\label{r2q}
  \tilde U_{l}(k',k) = \frac{2}{\pi}
  \int_0^\infty r'^2dr' 
  \int_0^\infty r^2dr\,
  j_l(k'r') U_{l}(r',r) j_l(kr)\;.
\end{equation}
In order to compare \emph{NA} potentials on a same footing we look 
for their reduced form.
Hence, starting with a potential in coordinate space
we transform it to momentum space using Eq.~(\ref{r2q}),
to then apply a momentum cutoff and transform it back to coordinate space.
The transformed potential $U_\Lambda$ is then used to
obtain its scattering observables solving Eq.~(\ref{wave}).

The presence of local contributions $V_l(r)$ in the hadronic part of
the \emph{NA} potential deserves special attention in the procedure
outlined above.
This feature takes place in the NSM approach, 
where the Hartree--Fock term is local. 
This also occurs for the direct term in coordinate--space folding 
models based on local effective interactions \cite{Amos2000}.
In such cases their momentum--space representation becomes
\begin{equation}
\label{vlocal}
  \tilde V_{l}(k',k) = \frac{2}{\pi}
  \int_0^\infty r^2dr\,
  j_l(k'r) V_{l}(r) j_l(kr)\;.
\end{equation}
What is interesting in this case is that its transformation
back to coordinate space, after the application of a cutoff at $\Lambda$,
result into a nonlocal potential $V_\Lambda(r',r)$. 
Hence, to the extent there is no sensitivity to the cutoff,
local potentials also have nonlocal equivalents.
This trend is somewhat observed in panels \emph{a'} of Figs. 
\ref{urra} and \ref{urrn2}, 
where the sharp diagonal structure for the imaginary part of the potential
(resembling locality) 
loses its sharpness as the cutoff is diminished.
For $\Lambda=4$~fm$^{-1}$ the equivalent potential is clearly nonlocal.

\begin{figure}
 \resizebox{0.9\textwidth}{!}{%
\includegraphics[angle=-0,origin=c,clip=true]{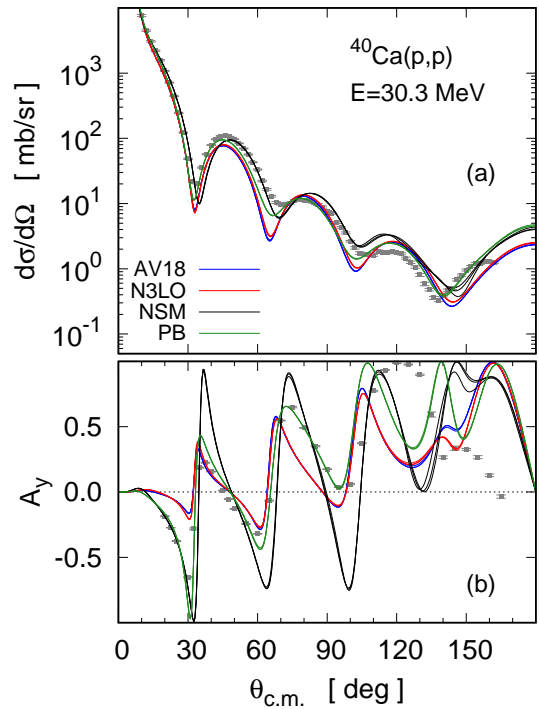}
 }
\vspace{-16mm}
\caption{{\protect\small
\label{xay030}     
  Differential cross section (a) and analyzing power (b) as functions of 
  the center--of--mass scattering angle for $^{40}$Ca($p$,$p$) scattering 
  at 30.3~MeV. 
  Overlapping solid curves correspond to results based on
  $\Lambda=6$, 4 and 2.6~fm$^{-1}$.
  Data from Ref. \cite{Ridley1964,Hnizdo1971}.
}}
\end{figure}

In Fig. \ref{xay030} we plot calculated differential cross section 
$d\sigma/d\Omega$ (a) and analyzing power $A_y$ (b) as functions of 
the center--of--mass scattering angle $\theta_{c.m.}$, 
for proton scattering off $^{40}$Ca at 30.3~MeV beam energy.
The data are taken from Refs. \cite{Ridley1964,Hnizdo1971},
included here in order to visualize the level of agreement of
each approach.
Blue and red curves denote results based on
AV18 and N3LO bare interactions, respectively.
Black curves represent results based on the NSM microscopic approach
using Gogny's effective interaction.
Green curves correspond to results based on the Perey--Buck--type
nonlocal parametrization of Ref. \cite{Tian2015}. 
In this figure we plot results with cutoffs at
$\Lambda=7$, 4 and 2.6~fm$^{-1}$, resulting in nearly full
overlap among curves of each case.
These results illustrate the level at which different cutoffs
above $\Lambda_A$ result in the same observables.
The differences among the four approaches indicate the extent
to which their corresponding reduced  potentials differ from each other. 

\begin{figure}[!ht]
\vskip +3mm     
 \resizebox{1.0\linewidth}{!}{%
\includegraphics[angle=-90,origin=c,clip=true]{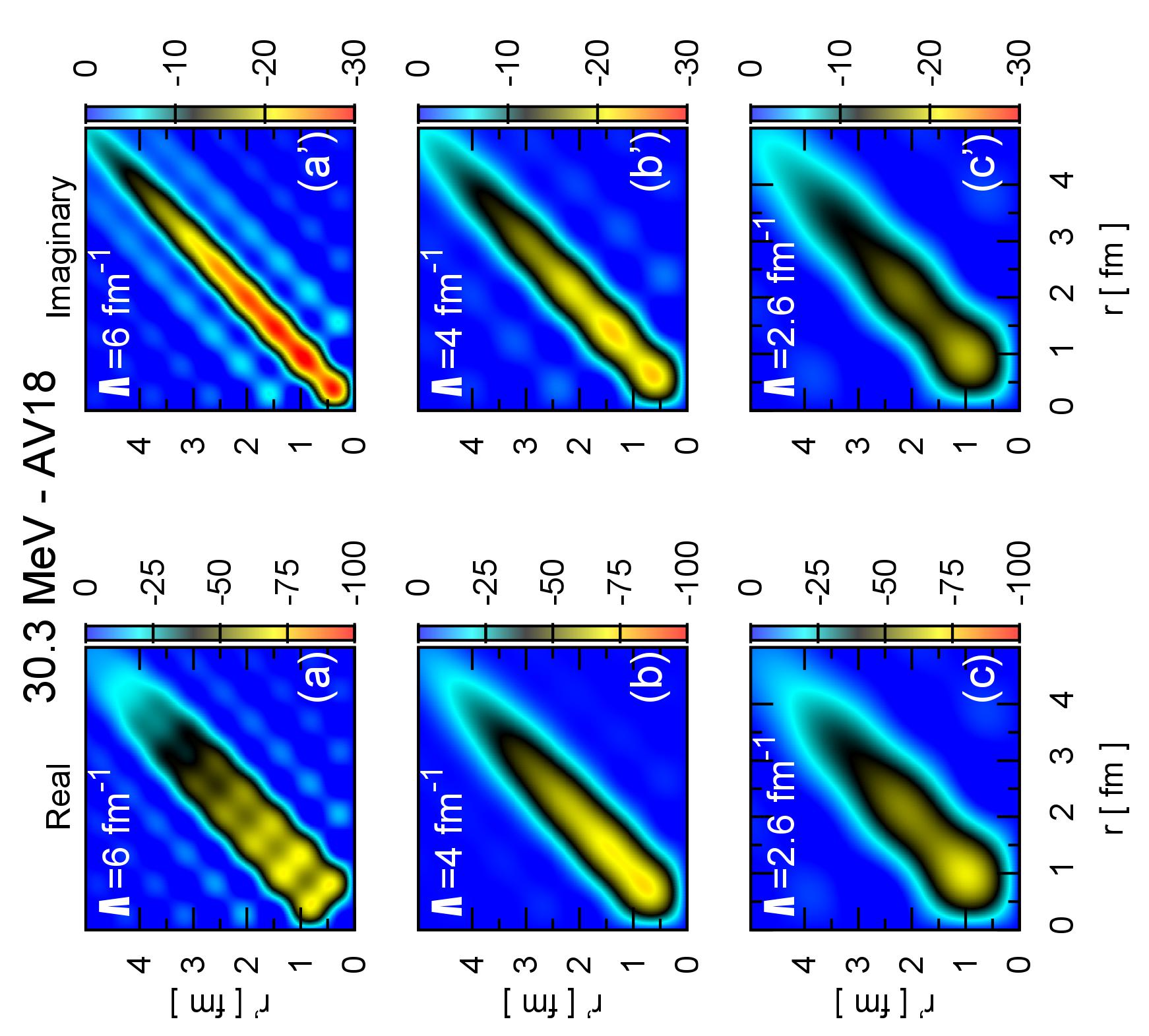}
 }
\vspace{-5mm}
\caption{{\protect\small
\label{urr030AV18}     
  Contour plots for the real (l.h.s. panels) and
  imaginary (r.h.s. panels) $s$--wave central optical
  potential as functions of the relative distance $r$ and $r'$.
  Potential for $p+^{40}$Ca scattering at 30.3~MeV.
  Results based on AV18 bare interaction.
  Frames a, b and c represent results
  based on $\Lambda=6$, $4$ and 2.6~fm$^{-1}$, respectively.
  All three potentials yield the same scattering observables.
  Color bar in units of MeV~fm$^{-1}$.
}}
\vspace{+5mm}
 \resizebox{1.0\linewidth}{!}{%
\includegraphics[angle=-90,origin=c,clip=true]{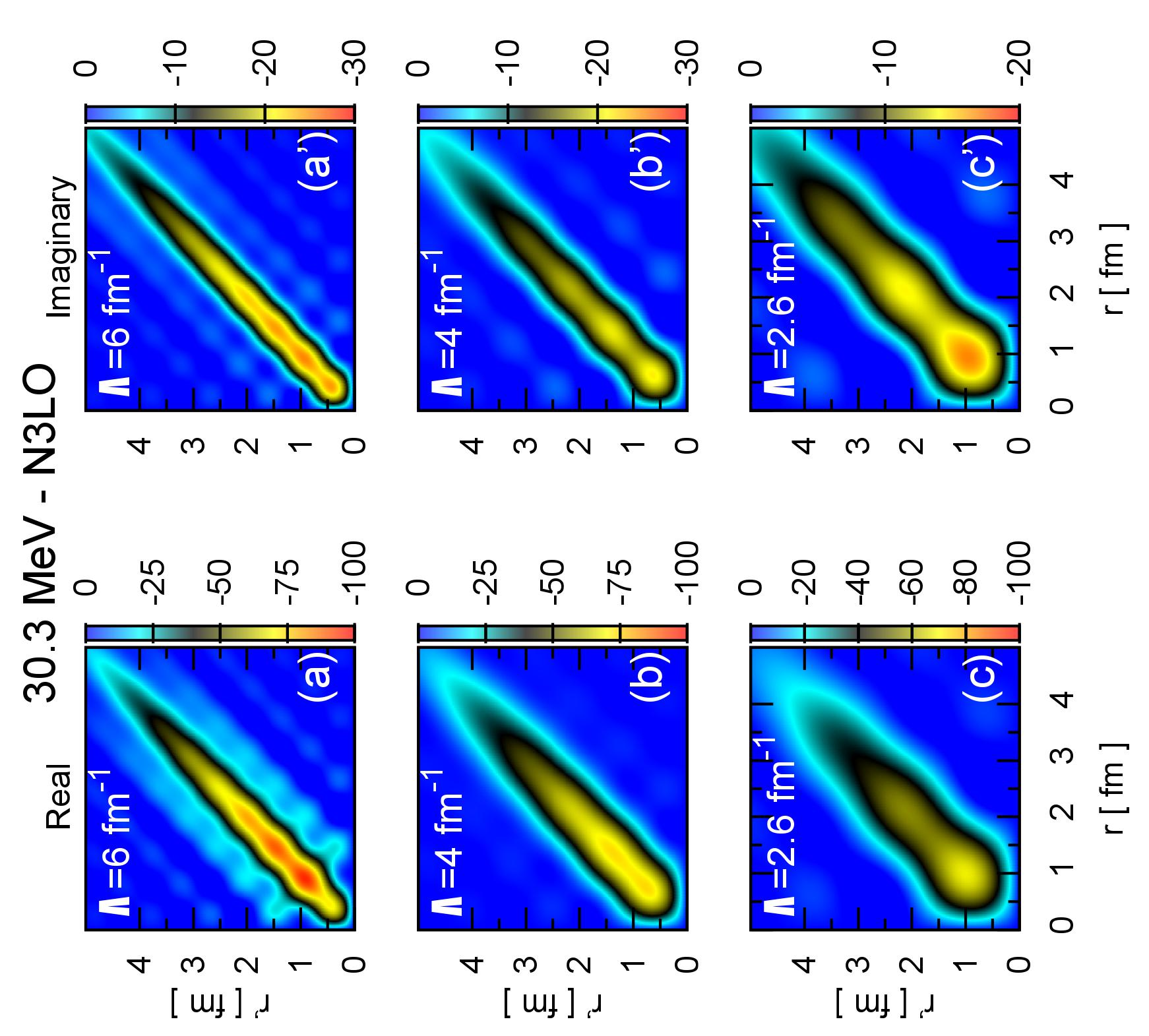}
 }
\vspace{-5mm}
\caption{{\protect\small
\label{urr030N3LO}     
  The same as Fig. \ref{urr030AV18}, but for N3LO bare interaction.
}}
\end{figure}

\begin{figure}[!ht]
\vskip +3mm     
 \resizebox{1.0\linewidth}{!}{%
\includegraphics[angle=-90,origin=c,clip=true]{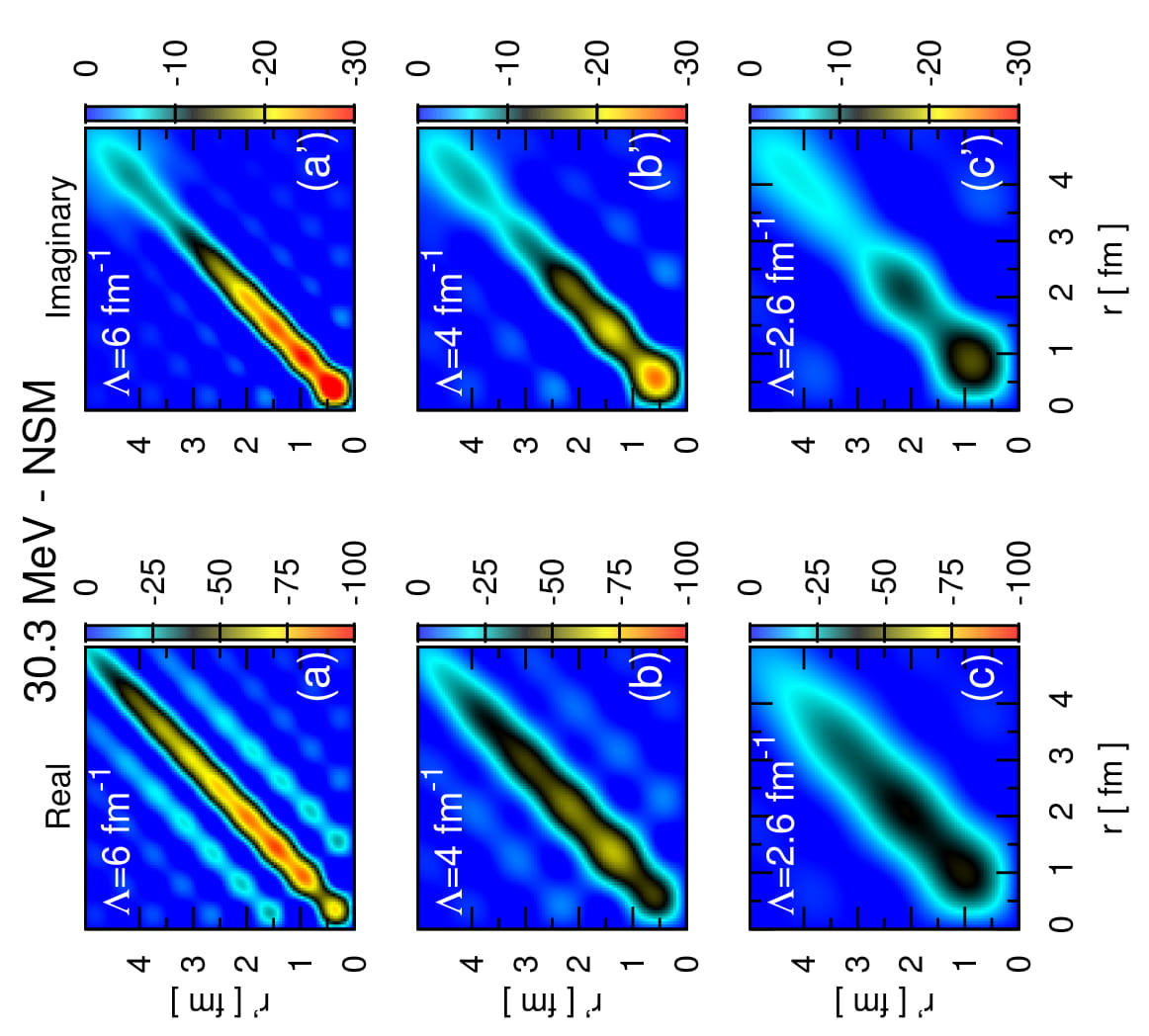}
 }
\vspace{-5mm}
\caption{{\protect\small
\label{urr030NSM}     
  The same as Fig. \ref{urr030AV18}, but based on NSM approach.
}}
\vspace{+23.5mm}
 \resizebox{1.0\linewidth}{!}{%
\includegraphics[angle=-90,origin=c,clip=true]{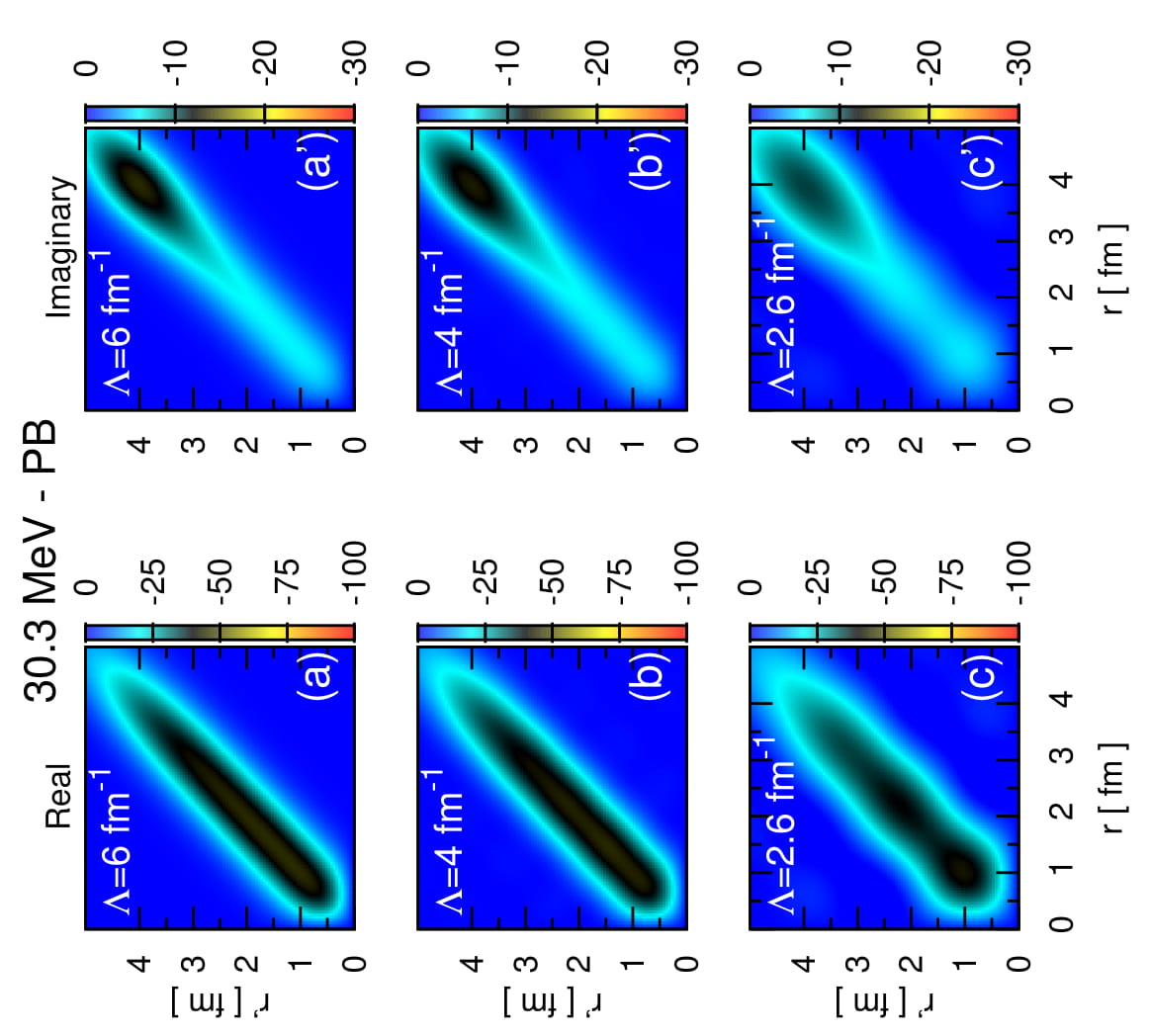}
 }
\vspace{-5mm}
\caption{{\protect\small
\label{urr030PB}     
  The same as Fig. \ref{urr030AV18}, but based on Perey--Buck--type
nonlocal parametrization of Ref.~\cite{Tian2015}.
}}
\end{figure}

In Figs. \ref{urr030AV18}--\ref{urr030PB} 
we show contour plots in the $(r,r')$ plane for the calculated 
$s$--wave $r'U(r',r)r$.
Panels \emph{a}, \emph{b} and \emph{c} correspond 
to $\Lambda=6$, 4 and 2.6~fm$^{-1}$,
respectively.
All cases meet the criteria $\Lambda\geq\Lambda_A$, resulting in
different nonlocal structure but sharing the same scattering observables.
On each Fig. panels on the l.h.s. (r.h.s.) correspond to their real
(imaginary) parts.
Plots in Figs. \ref{urr030AV18} and \ref{urr030N3LO} are based
on AV18 and N3LO bare potentials, respectively.
Plots in Fig. \ref{urr030NSM} are based on NSM microscopic approach,
while Fig. \ref{urr030PB} is based on Perey--Buck--type parametrization
of Ref. \cite{Tian2015}.
Note that the scale of panels \emph{a} in 
Figs. \ref{urr030AV18}--\ref{urr030PB} range from $-100$~MeV up to zero. 
In order to make comparable these Figs. all corresponding 
panels use the same scale. 
Panels \emph{c} show the reduced potential since
the cutoff applied in these cases corresponds to $\Lambda_A$ given by
Eq. (\ref{law}).
We observe the following: \\

\noindent\emph{a.--}
All upper frames \emph{a} and \emph{a'} show the nonlocal 
behavior of the potential closest to its original form. 
By closest we mean they all have been already affected by a cutoff,
though relatively large with respect to the critical one.
Observing panel \emph{a} for AV18 we identify a rather flat diagonal band 
of about $\sim$1~fm width, with strength nearing $-75$~fm.
This band, which does not appear in the other cases, 
weakens near the surface ($r\approx r'\approx 4$~fm).
Frames \emph{a} for N3LO and NSM exhibit sharper potentials along the 
diagonal, vanishing around 4~fm.  
The corresponding panel for PB shows a weaker and smooth potential with
an elongated oval shape.
With the exception of Fig. \ref{urr030PB} for PB, 
all potentials exhibit fine structures in the form of regular spots 
away from the diagonal. \\

\noindent\emph{b.--}
As the cutoff $\Lambda$ diminishes,
the coordinate--space potential becomes less structured.
This is observed in all four cases.
In this regard note how similar is the reduced potential
based on AV18 with that based on the N3LO bare interaction. 
In the case of NSM in Fig.~\ref{urr030NSM} 
the real part of the reduced potential is more shallow at the center 
than in the previous ones, also weakening as the surface is reached at
$r\sim r'\approx 2$~fm.
In the case of PB, the real part resembles very much the ones 
based on NSM.\\

\noindent\emph{c.--}
The imaginary part of the potentials shown at the r.h.s. of each Fig. 
based on AV18 and N3LO exhibits minor differences, 
both in their raw and reduced forms.
The other two cases, namely NSM and PB are much different. 
The former starts at frame \emph{a'} with a narrow band and end up
at panel \emph{c'} with a less structures absorptive component confined 
to radii below 2~fm.
The opposite occurs for PB,
where the absorptive part takes place above radii of 3~fm, remaining
in frame \emph{c'} away from the bulk of the nucleus.

\subsection{Reduced potentials in coordinate space}
\begin{figure}
 \resizebox{1.0\textwidth}{!}{%
\includegraphics[angle=-0,origin=c,clip=true]{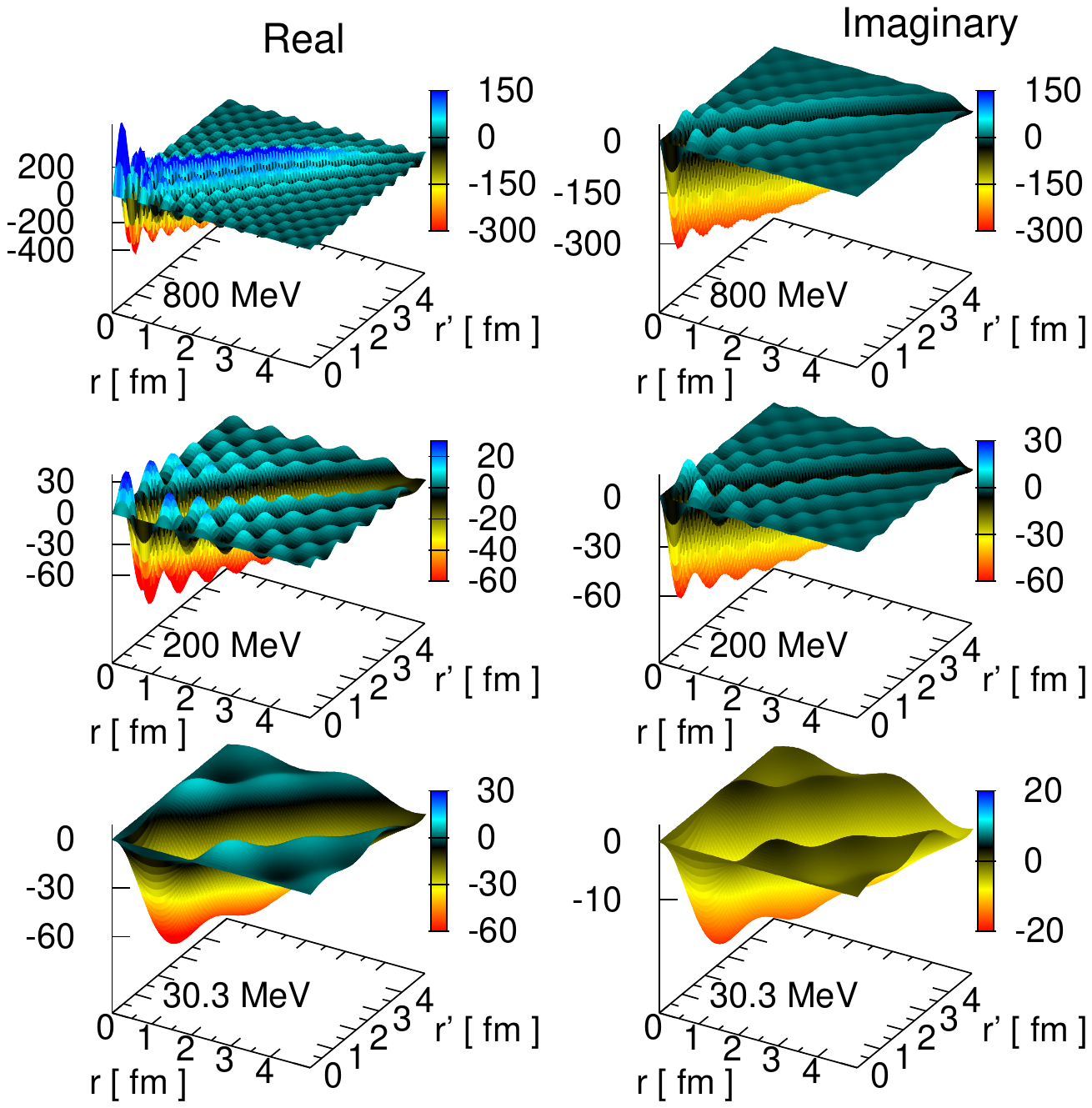}
 }
\vspace{-10mm}
\caption{{\protect\small
\label{RRsurfaces}     
  Surface plots of real (l.h.s. graphs) and
  imaginary (r.h.s. graphs) $s$--wave reduced central optical 
  potential 
  $rU(r,r')r'$ as functions of the relative distance $r$ and $r'$. 
  Potential for $p+^{40}$Ca scattering at 800, 200 and 30.3~MeV.
  Results based on AV18 bare interaction.
  Color bars and vertical axes are in units of MeV~fm$^{-1}$.
}}
\end{figure}

As discussed in the previous section,
the critical cutoff $\Lambda_A$ lead to 
optical potentials with least structure in coordinate space.
This characteristic is observed in all partial waves as well as energies.
In order to illustrate this feature in 3D surface
plots, we consider $s$--wave potentials (central part) for
$p+^{40}$Ca elastic scattering at 800, 200 and 30.3~MeV.
In Fig.~\ref{RRsurfaces} we show surface plots of the real (l.h.s. graphs) and
  imaginary (r.h.s. graphs) $s$--wave reduced central optical
  potential $rU(r,r')r'$, as functions of $r$ and $r'$.
These results are based on AV18 bare interaction.
The critical cutoffs for each case are 8, 4, and 2.6~fm$^{-1}$,
respectively.
Color bars and vertical axes are in units of MeV~fm$^{-1}$.
As observed, all potentials exhibit pronounced structures in both
real and imaginary components. 
The case of 30.3~MeV appears the most smooth of all.
As the energy increases, the potential features strong oscillating patterns
near and away of the diagonal.
We have to stress that all these potentials
are the least structured in coordinate space, consistent with
the scattering observables they have in their original form, namely
with large cutoff.

\begin{figure}
 \resizebox{0.9\textwidth}{!}{%
\includegraphics[angle=-0,origin=c,clip=true]{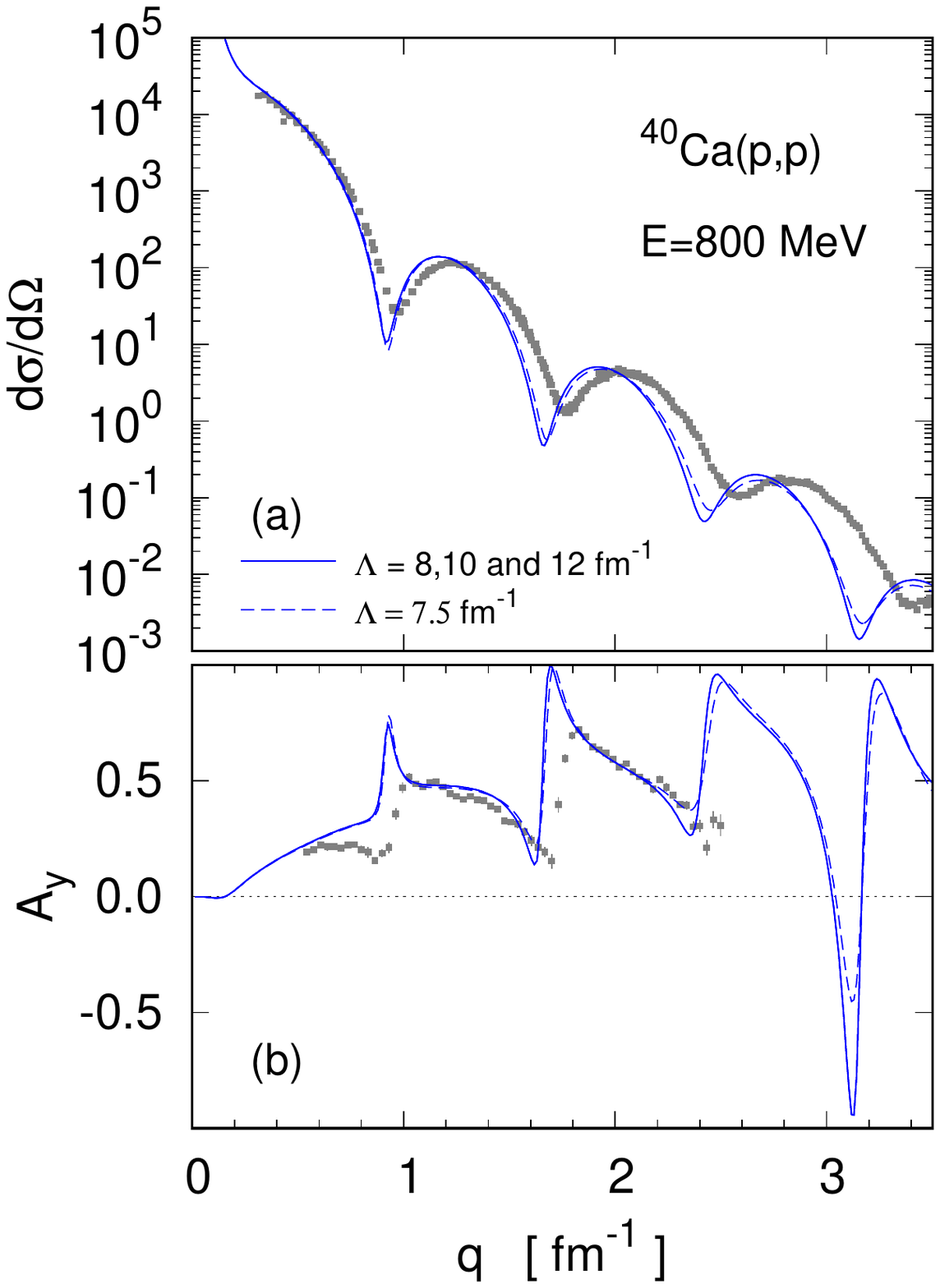}
 }
\vspace{-15mm}
\caption{{\protect\small
\label{xay800}     
  Differential cross section (a) and analyzing power (b) as functions of 
  the momentum transfer for $^{40}$Ca($p$,$p$) scattering at 800~MeV. 
  Overlapping solid curves correspond to results based on
  $\Lambda=12$, 10 and 8~fm$^{-1}$.
  Dashed curves correspond to $\Lambda=7.5$~fm$^{-1}$.
  Data taken from Refs. \cite{Ray1981,Igo1979}
}}
\end{figure}

For completeness in this discussion, in Fig.~\ref{xay800} we
plot the differential cross section (a) and analyzing power (b) 
as functions of the momentum transfer for $^{40}$Ca($p$,$p$) 
scattering at 800~MeV.
Here overlapping solid curves correspond to results based on
$\Lambda=12$, 10 and 8~fm$^{-1}$. 
Dashed curved correspond to $\Lambda=7.5$~fm$^{-1}$, below the
critical value.
The data are taken from Refs. \cite{Ray1981,Igo1979}.
As observed, this case exhibits the same features as the ones discussed
in the previous sections, namely there is range of cutoffs above
$\Lambda_A$ which do not alter the calculated scattering observables.
The case $\Lambda=7.5$~fm$^{-1}$, below $\Lambda_A$, displays clearly 
its differences with the actual observables.
The specific issue on the level of agreement of the model with the data 
requires a more dedicated study which is beyond the scope of this work.
Regardless on any change at the level of the microscopic description
at this energy, it is safe to state that the nonlocal structure of
the calculated potential will remain essentially the same as the one
shown in Fig.~\ref{RRsurfaces}.

\section{Summary and conclusions}
\label{summary}
  We have investigated the nonlocal structure of optical model potentials
  for \emph{NA} based on microscopic approaches. 
  To this purpose, \emph{in-medium} folding optical potentials
  have been calculated in momentum space and their corresponding
  coordinate-space counterpart have been examined, 
  focusing our attention on their nonlocal shape. 
  The \emph{NN} effective interaction consists of the actual 
  full off--shell $g$ matrix in Brueckner--Hartree--Fock approximation. 
  The nonlocality of effective interactions is preserved throughout all
  the stages of the the calculation. 
  The bare interactions under consideration are Argonne $v_{18}$ potential 
  and chiral next--to--next--to--next--to--leading order bare interaction.
  The study has been focused on proton elastic scattering 
  off $^{40}$Ca at beam energies between 30 and 800~MeV.
  Applications at 30.3~MeV have also been studied using the microscopic 
  NSM optical potential as well as phenomenological Perey--Buck type.
  We have found that the gradual suppression of high--momentum 
  contributions of the optical potential results in quite 
  different--looking coordinate--space counterparts.
  Despite this non-uniqueness in their nonlocal structure, the implied 
  scattering observables remain unchanged for momentum cutoff above a
  critical one, which depends on incident energy of the projectile. 

Folding optical model potentials based on any realistic \emph{NN} 
interaction, as the one used in this work, 
offer the possibility to explore in a systematic way 
nucleon collisions at energies ranging from tens of MeV up to
near GeV energies. 
A crucial point in the choice of this framework is the fact that 
all sources of nonlocalities are preserved throughout, namely
the determination of the effective \emph{NN} interaction 
and also the evaluation of the potential itself. 
Thus, momentum--space \emph{in-medium} folding potentials constitute 
a general microscopic starting point to investigate the optical model 
and its nonlocal features. 
In this framework the potential is evaluated in momentum space,
representation which enables the treatment of nonlocalities in a natural
way. In this regard the model is not a particular one,
but instead a general one to account for these features.

  The identification of an energy--dependent critical cutoff 
  that guarantees invariability in the scattering observables is an
  important finding in this work, particularly due to its implications
  on the coordinate--space structure of equivalent potentials. 
  We have demonstrated that the nonlocal shape of the potential becomes 
  less structured as the cutoff is decreased until a critical value,
  which guarantees that scattering observables remain unchanged.  
  An immediate implication of this result is that any attempt to quantify 
  the degree of nonlocality of a given potential is meaningless.
The non uniqueness of nonlocality in coordinate space
also questions the systematic resort to Gaussian nonlocalities 
in the design of phenomenological potentials,
as first introduced by Perey and Buck in the 60s. 
Despite this limitation, we have found that potentials constructed
in any representation, even local ones in coordinate space, 
can be subject to momentum cutoff which render the same 
scattering observables. The resulting potential in coordinate space
is always nonlocal, although less structured when the cutoff is at
$\Lambda_A$. 
This feature may shed light in works aiming to
generate phenomenological nonlocal optical potentials.

From a practical point of view, the identification of the 
energy--dependence of the critical cutoff $\Lambda_A$ allows 
to identify beforehand the
range of momenta relevant in the evaluation of an optical potential 
in momentum space. 
From the present study we can safely state that momentum components 
of the optical potential above $\Lambda_A$ are irrelevant.
The generalization of this empirical law for $\Lambda_A$ 
to any mass number $A$ in the nuclear chart constitutes an 
extension of this work.

\begin{acknowledgements}
H.F.A. thanks colleagues at CEA, Bruy\`eres-le-Ch\^atel, France, for their 
hospitality during his stay where part of this work was done.
\end{acknowledgements}
%
%
 \bibliography{misreferencias}
 \end{document}